\documentclass[3p,preprint,times,superscriptaddress]{elsarticle}
\usepackage{amssymb}
\usepackage{verbatim}
\usepackage{slashed}
\usepackage{epsfig}
\usepackage{graphicx}
\usepackage{subfigure}
\usepackage{amsmath}
\usepackage{mathrsfs}
\usepackage{bm}

\usepackage[figuresright]{rotating}

\journal{elsevier: UB-ECM-PF-70/12, ICCUB-12-097}

\begin{document}

\begin{frontmatter}





\title{Investigating the heavy quarkonium radiative transitions with the effective Lagrangian method}


\author[UB]{Zhi-Guo He}
\ead{hzgzlh@gmail.com}

\address[UB]{\small{\it{Departament d'Estructura i Constituents de
la Mat\`eria
                  and Institut de Ci\`encies del Cosmos}}\\
        \small{\it{Universitat de Barcelona}}\\
        \small{\it{Diagonal, 647, E-08028 Barcelona, Catalonia, Spain.}}}

\begin{abstract}
In this work, we study the radiative decay of heavy quarkonium states by using the effective
Lagrangian approach. Firstly, we construct the spin-breaking terms in the effective Lagrangian
for the $nP\leftrightarrow mS$ transitions and determine the some of the coupling constants by
fitting the experimental data. Our results show that in $\chi_{cJ}$, $\psi(2S)$, $\Upsilon(2S)$,
and $\Upsilon(3S)$ radiative decays, the spin-breaking effect is so small that can be ignored.
Secondly, we investigate the radiative decay widths of the $c\bar{c}(1D)$ states and find the
if $\psi(3770)$ is a pure $^3D_1$ state its radiative decay into $\chi_{cJ}+\gamma$ roughly
preserve the heavy-quark spin symmetry, while if it is a $S-D$ mixing state with mixing angel
$12^{\circ}$ the heavy quark-spin symmetry in its radiative decay and in the radiative decay of
$\psi(3686)$ will be largely violated. In the end, we show that combining the radiative decay
and the light hadron decay of $P$-wave $\chi_{bJ}(1,2P)$ can provide another way to extract the
information of the color-octet matrix element in the context of non-relativistic QCD (NRQCD)
effective theory, and our result is consistent with potential NRQCD hypothesis.

\end{abstract}


\end{frontmatter}


\section{Introduction}\label{sec1}

The heavy quarkonium states that are constituted by heavy-quark ($Q$) and
anti-quark ($\bar{Q}$) pair provide an ideal laboratory to study the dynamics
of strong interaction from both perturbative and non-perturbative aspects.
In recent years, thanks to the large amount data accumulated in electron-positron
colliders and hadronic colliders, more accurate and new properties of them
have been obtained. Especially, many new resonances were discovered, which gave
rise to a great renewed theoretical interest in studying their spectra and decays
(for recent reviews see Ref.\cite{Swanson:2006st} and references therein).

For the states below open flavor threshold ($D\bar{D}$ for charmonium and
$B\bar{B}$ for bottomonium), they have relative narrow width because they
can not decay through the Okubo-Zweig-Iizuka allowed decay mechanism. Their
radiative decay width could reach hundred kev level, therefore, contributes
a considerable branching ratio. On the experimental side, the radiative
transitions among heavy quarkonia also play an important role in searching
for the new states. Theoretically, the heavy quarkonium states are approximate
nonrelativistic systems. Their annihilation decays are sensitive to the wave
functions of $Q\bar{Q}$ at small distance, on the contrary the radiative decay can help us
to probe the behavior of wave functions at long-distance.  Besides the intrinsic
scale $\Lambda_{\rm{QCD}}$, heavy quarkonia are characterized by a hierarchy of
three energy scales, $m_{Q}$ the heavy quark mass, $m_{Q}v_{Q}$ and $m_{Q}v_{Q}^2$
the typical momentum and energy of the heavy quark, where $v_{Q}\ll1$ is velocity
of the heavy quark in the rest frame of the heavy meson. The nonrelativistic
effective field theories, nonrelativistic QCD (NRQCD) \cite{Caswell:1985ui,Thacker:1990bm,Bodwin:1994jh},
and potential NRQCD (pNRQCD) \cite{Pineda:1997bj,Brambilla:1999xf,Brambilla:2004jw}
are suitable tools to separate the physics in different energy scales. Recently,
the magnetic dipole ($M1$) transition as well as the radiative decay of $X(3872)$
was studied within the framework of pNRQCD\cite{Brambilla:2005zw}. Besides from the
model-independent perspective, the radiative transitions among heavy quarkonia have
already been extensively studied within potential model approach (here we refer
Ref.\cite{Eichten:2007qx} as a comprehensive review).


In this letter, we will not only calculate the radiative decay widths, but will also
do some further analysis by taking into account the spin-breaking effect or the $S-D$ mixing
effect in the radiative decays. We will also combine the radiative decay with the light
hadron (LH) decay of the $\chi_{bJ}(nP)$ states to extract the information of the color-octet
(CO) matrix elements in NRQCD\cite{Bodwin:1994jh} and pNRQCD\cite{Brambilla:2001xy,Brambilla:2002nu}.
In this work, we plan to employ the effective Lagrangian approach, which can exploit
the heavy quark spin symmetry order by order through the expansion of $1/m_{Q}$.
The rest of this letter is organized as follows. A brief description of the effective
Lagrangian approach will be given in Section 2, and then is used to study the
$S\leftrightarrow P$ transitions in $c\bar{c}$ and $b\bar{b}$ systems by taking into account
spin-breaking effect. In section 3, we will study the $\psi(^3D_J)\to \chi_{cJ^{\prime}}+\gamma$
transitions. In section 4, we will relate the transitions of $\chi_{bJ}(nP)\to\Upsilon(nS)+\gamma$
to the LH decay of $\chi_{bJ}(nP)$ states to determine the ratios of the CO matrix
elements $m_b^2\mathcal{H}_{8}(nP)$ to the corresponding color-singlet (CS) matrix
elements $\mathcal{H}_{1}(nP)$, where $\mathcal{H}_{1}(nP)=\langle\chi_{bJ}(nP)|\mathcal{O}_{1}(^3P_{J})|\chi_{bJ}(nP)\rangle$,
and $\mathcal{H}_{8}(nP)=\langle\chi_{bJ}(nP)|\mathcal{O}_{8}(^3S_{1})|\chi_{bJ}(nP)\rangle$
\cite{Bodwin:1994jh}. A short summary and conclusion will be presented in the last section.

\section{Effective Lagrangian For Radiative Transitions}

The heavy charmonium states can be classified according to the spectroscopic notation
$n^{2S+1}L_{J}$ , where $n=1,2,\ldots$ is the radial quantum number, $S=0,1$ is the
total spin of the heavy quark pair, $L=0,1,2\ldots$ (or $S,P,D\ldots$) is the orbital
angular momentum, and $J$ is the total angular momentum. They have parity $P=(-1)^{L+1}$
and charge conjugation $C=(-1)^{L+S}$. As mentioned in the introduction, NRQCD and pNRQCD
are a good starting point to describe this system. The LO NRQCD Lagrangian is invariant under
S=$SU(2)_Q\otimes SU(2)_{\bar Q}$ spin symmetry group, an approximate symmetry of the heavy
quarkonium states, that is inherited in the subsequent effective theories. Hence, it is most
convenient to introduce hadronic spin-symmetry multiplets, in an analogous way as it was
initially done in Heavy Quark Effective Theory (HQET) \cite{Neubert:1993mb}.

For heavy quarkonium states, this formalism was developed in Ref. \cite{Casalbuoni:1992yd}.
The states have the same radial number $n$ and the same orbital momentum $L$ can also be
expressed by means of a single multiplet:
$J^{\mu_1\ldots\mu_L}$
\cite{Casalbuoni:1992yd},
\begin{eqnarray}
J^{\mu_1\ldots\mu_L}&=&\frac{1+\slashed{v}}{2}(H_{L+1}^{\mu_1\ldots\mu_L\alpha}\gamma_{\alpha}+\frac{1}{\sqrt{L(L+1)}}
\sum_{i=1}^{L}\epsilon^{\mu_i\alpha\beta\gamma}v_{\alpha}\gamma_{\beta}H_{L\gamma}^{\mu_1\ldots\mu_{i-1}\mu_{i+1}\ldots\mu_L}
\nonumber\\&+&
\frac{1}{L}\sqrt{\frac{2L-1}{2L+1}}\sum_{i=1}^{L}(\gamma^{\mu_i}-v^{\mu_i})H_{L-1}^{\mu_1\ldots\mu_{i-1}\mu_{i+1}\ldots\mu_L}
\nonumber\\&-&
\frac{2}{L\sqrt{(2L-1)(2L+1)}}\sum_{i<j}(g^{\mu_i\mu_j}-v^{\mu_i}v^{\mu_j})\gamma_{\alpha}
H_{L-1}^{\alpha\mu_1\ldots\mu_{i-1}\mu_{i+1}\ldots\mu_{j-1}\mu_{j+1}\ldots\mu_L}
\nonumber\\&+&K_{L}^{\mu_1\ldots\mu_{L}}\gamma^{5})\frac{1-\slashed{v}}{2}
\end{eqnarray}
where $v^{\mu}$ is the four-velocity associated to the multiplet
$J^{\mu_1\ldots\mu_L}$ (not to be mistaken by $v_{Q}$, the typical
velocity of the heavy quark in the heavy quarkonium rest frame),
$K_{L}^{\mu_1\ldots\mu_{L}}$ represents the spin-singlet effective
field, and
$H_{L-1}^{\mu_1\ldots\mu_{L-1}}$,$H_{L}^{\mu_1\ldots\mu_{L}}$ and
$H_{L+1}^{\mu_1\ldots\mu_{L+1}}$ represent the three spin-triplet
effective fields with $J=L-1,L$, and $L+1$ respectively. The four
tensors are all completely symmetric and traceless and satisfy the
transverse condition
\begin{equation}\label{trans}
v_{\mu_i} K_{L}^{\mu_1\ldots\mu_i\ldots\mu_{L}}=0\quad ,\quad  v_{\mu_j} H_{J}^{\mu_1\ldots\mu_j\ldots\mu_{J}}=0.
\end{equation}
$i=1,\dots , L$, $j=1,\dots , J$. The properties of $H$ and $K$ under
parity, charge conjugation and heavy quark spin transformations can be
easily obtained by assuming that the corresponding transformation rules
of the multiplet $J^{\mu_1\ldots\mu_L}$ follow as:
\begin{subequations}
\begin{eqnarray}
J^{\mu_1\ldots\mu_L}\stackrel{P}{\longrightarrow}\gamma^{0}J_{\mu_1\ldots\mu_L}\gamma^{0},
v^{\mu}\stackrel{P}{\longrightarrow}v_{\mu},
\end{eqnarray}
\begin{eqnarray}
J^{\mu_1\ldots\mu_L}\stackrel{C}{\longrightarrow}(-1)^{L+1}C[J_{\mu_1\ldots\mu_L}]^{T}C,
\end{eqnarray}
\begin{eqnarray}
J^{\mu_1\ldots\mu_L}\stackrel{\rm S}{\longrightarrow}SJ_{\mu_1\ldots\mu_L}S^{\prime\dagger},
\end{eqnarray}
\end{subequations}
where $C$ is the charge conjugation matrix ($C=i\gamma^{2}\gamma^{0}$
in the Dirac representation), and $S\in SU(2)_Q$ and $S^{\prime} \in SU(2)_{\bar Q}$
correspond to the heavy quark and heavy antiquark spin symmetry groups
($[S,\slashed{v}]=[S^{\prime},\slashed{v}]=0$).

Since we are going to consider the $S$, $P$, and $D$ wave states radiative
decay,it will be helpful to give the explicit expressions of the $S$-, $P$-,
and $D-$wave multiplets that follow from Eq.(1). For the $L=S$ case, we have
\begin{equation}
J=\frac{1+\slashed{v}}{2}(H_1^{\mu}\gamma_{\mu}-K_0\gamma^{5})\frac{1-\slashed{v}}{2},
\end{equation}
for the $L=P$ case,
\begin{equation}
J^{\mu }=\frac{1+\slashed{v}}{2}\Big\{
H_2^{\mu \alpha } \gamma_\alpha  + {1 \over \sqrt{2}}
\epsilon^{\mu \alpha \beta \gamma} v_\alpha \gamma_\beta H_{1
\gamma}+\frac{1}{\sqrt{3}}
(\gamma^{\mu} -v^{\mu}) H_0  + K_1^{\mu }\gamma_5 \Big\}\frac{1-\slashed{v}}{2},
\end{equation}
and for $L=D$ case,
\begin{eqnarray}
&J^{\mu\nu}=\frac{1+\slashed{v}}{2}\Big\{H_{3}^{\mu \nu \alpha}\gamma_{\alpha}+
\frac{1}{\sqrt{6}}(\epsilon^{\mu \alpha \beta \gamma} v_{\alpha}\gamma_{\beta}H^{\nu}_{2\gamma}+
\epsilon^{\nu\alpha \beta \gamma} v_{\alpha}\gamma_{\beta}H^{\mu}_{2\gamma})&
\nonumber\\
&+\sqrt{\frac{3}{20}}[(\gamma^{\mu}-v^{\mu})H_{1}^{\nu}+(\gamma^{\nu}-v^{\nu})H_{1}^{\mu}
-\frac{2}{3}(g^{\mu \nu}-v^{\mu}v^{\nu})\gamma_{\alpha}H_{1}^{\alpha}]+K_{2}^{\mu\nu}\gamma^{5}\Big\}\frac{1-\slashed{v}}{2}&
\end{eqnarray}

At the leading order of $1/m_{Q}$ expansion, the radiative transitions between $mS$
and $nP$ states, and between $mP$ and $nD$ states can be described by the Lagrangian
given in Ref. \cite{Casalbuoni:1992yd,DeFazio:2008xq}:
\begin{subequations}\label{vertex}
\begin{equation}\label{vertexPS}
\mathcal{L}^{SP}=\sum_{m,n}\delta_{Q}^{nP,mS}\mathrm{Tr}[\bar{J}(mS)J_{\mu}(nP)]
v_{\nu}F^{\mu\nu}+\mathrm{h.c.}\;,
\end{equation}
\begin{equation}\label{vertexPD}
\mathcal{L}^{PD}=\sum_{m,n}\delta_{Q}^{nD,mP}\mathrm{Tr}[\bar{J}^{\alpha}(mP)J_{\alpha\mu}(nD)]
v_{\nu}F^{\mu\nu}+\mathrm{h.c.}\;,
\end{equation}
\end{subequations}
where $\delta_{Q}^{nP,mS}(Q=c,b)$ and $\delta^{mP,nD}(Q=c,b)$ are the coupling
constants, and $F^{\mu\nu}$ is the electromagnetic tensor. The Lagrangian in
Eq. (\ref{vertex}) preserves parity, charge conjugation, gauge invariance and
heavy quark and antiquark spin symmetry.

The radiative decays of the low lying $S$- and $P$- wave states have been well measured.
It will be interesting to do some delicate analysis beyond leading order.
One important higher order contribution comes from the spin-breaking effect which
is due to the spin-spin $\mathbf{S}_{Q}\cdot\mathbf{S}_{\bar{Q}}$, spin-orbit $\mathbf{L}\cdot\mathbf{S}$,
and tensor $\frac{(\mathbf{S}_{Q}\cdot \mathbf{r})(\mathbf{S}_{\bar{Q}}\cdot \mathbf{r})}{r^2}
-\frac{\mathbf{S}_{Q}\cdot\mathbf{S}_{\bar{Q}}}{3}$ interactions in pNRQCD \cite{Pineda:2000sz}
(or in potential models), where $\mathbf{S}_{Q}$ and $\mathbf{S}_{\bar{Q}}$ are the spin
of the quark and anti-quark respectively, $\mathbf{S}=\mathbf{S}_{Q}+\mathbf{S}_{\bar{Q}}$,
and $\mathbf{L}$ is the orbital angular momentum of the heavy meson. To figure out all
the spin-breaking terms in the effective Lagrangian, it will be more perspicuous to
construct them in the rest frame of the heavy meson, where the $4\times4$ dimensional space
is reduced to the $2\times2$ dimensional space. In the two component notation, the field $J$
and $J^{\mu}$ is simplified as:
\begin{equation}
J=\vec{H}\cdot\vec{\sigma}+K_{0},\;J^{i}=(H_2^{ij}+\frac{1}{\sqrt{2}}\epsilon^{ijk}H^{k}_{1}
+\frac{\delta^{ij}}{\sqrt{3}}H_{0})\sigma^{j}+K_{1}^{i},
\end{equation}
where $\vec{\sigma}$ is the Pauli matrix. The Lagrangian in Eq.(\ref{vertexPS}) becomes:
\begin{equation}
\mathcal{L}^{SP}=\sum_{m,n}\delta_{Q}^{nP,mS}\mathrm{Tr}[J^{\dagger}(mS)
J^{i}(nP)]E^{i}+\mathrm{h.c.}\;.
\end{equation}
In the $2\times2$ dimensional space, the spin breaking terms can only be in the form
of $\vec{\alpha}\cdot\vec{\sigma}$, where $\vec{\alpha}$ is an arbitrary three dimensional
vector. After analyzing all the possible combinations of the field and $\vec{\sigma}$
operators, we find that there are three independent spin-breaking terms at sub-leading
order in $1/m_{Q}$, which are given by
\begin{subequations}\label{Spin-Breaking}
\begin{equation}
\mathcal{L}^{SP}_{QS}=\delta^{nP,mS}_{SS}(\mathrm{Tr}[J^{\dagger}\sigma^{j}J^{i}\sigma^{i}])E^{i}+\mathrm{h.c.}\;,
\end{equation}
\begin{equation}
\mathcal{L}^{SP}_{QL}=-i\frac{\delta^{nP,mS}_{LS}}{2}\epsilon^{ijk}(\mathrm{Tr}[J^{\dagger}\sigma^{j}J^{k}]
-\mathrm{Tr}[J^{\dagger}\sigma^{k}J^{j}])E^{i}+\mathrm{h.c.}\;,
\end{equation}
\begin{equation}
\mathcal{L}^{SP}_{QT}=\frac{\delta^{nP,mS}_{T}}{2}(\mathrm{Tr}[J^{\dagger}\sigma^{j}J^{j}\sigma^{i}]
+\mathrm{Tr}[\dagger\sigma^{i}J^{j}\sigma^{j}])E^{i}+\mathrm{h.c.}\;,
\end{equation}
\end{subequations}
where $\delta^{nP,mS}_{SS}$, $\delta^{nP,mS}_{LS}$ and $\delta^{nP,mS}_{T}$ are the coupling constants
that are suppressed by $1/m_{Q}^2$,since the spin-breaking potentials is of $\mathcal{O}(1/m_{Q}^2)$
compared to the static potential\cite{Pineda:2000sz}.

\begin{table}
\caption{The numerical values of the coupling constants $\delta_{QJ}^{nP,mS} (\mathrm{GeV}^{-1})$
determined by fitting the experimental decay widths.}
\vspace{0.8cm}
\hspace{-0.6cm}
\begin{tabular}{|c|c|c|c|}
     \hline
     \multicolumn{2}{|c|}{Charmonium} &\multicolumn{2}{|c|}{Bottomonium} \\
     \hline
     Decay Width (keV)& $\delta_{cJ}^{nP,mS}(\rm{GeV}^{-1})$& Decay Width (keV)&
     $\delta_{bJ}^{nP,mS}(\rm{GeV}^{-1})$\\
     \hline
     $\Gamma(\chi_{c0}\to J/\psi+\gamma)=121.7\pm10.9$& $(2.13\pm0.95)\times10^{-1}$&$\Gamma(\Upsilon(2S)\to \chi_{b0}(1P)+\gamma)=1.22\pm0.16$&$(9.01\pm0.50)\times10^{-2}$\\
     \hline
     $\Gamma(\chi_{c1}\to J/\psi+\gamma)=295.8\pm21.5$& $(2.31\pm0.08)\times10^{-1}$ &$\Gamma(\Upsilon(2S)\to \chi_{b1}(1P)+\gamma)=2.21\pm0.22$&$(9.89\pm0.50)\times10^{-2}$\\
     \hline
     $\Gamma(\chi_{c2}\to J/\psi+\gamma)=384.2\pm26.6$& $(2.29\pm0.08)\times10^{-1}$ &$\Gamma(\Upsilon(2S)\to \chi_{b2}(1P)+\gamma)=2.29\pm0.22$&$(9.86\pm0.47)\times10^{-2}$\\
     \hline
     $\Gamma(\psi^{\prime}\to \chi_{c0}+\gamma)=29.2\pm1.3$& $(2.25\pm0.04)\times10^{-1}$ &$\Gamma(\Upsilon(3S)\to \chi_{b0}(2P)+\gamma)=1.20\pm0.16$&$(1.39\pm0.09)\times10^{-1}$\\
     \hline
     $\Gamma(\psi^{\prime}\to \chi_{c1}+\gamma)=28.0\pm1.5$& $(2.36\pm0.06)\times10^{-1}$ &$\Gamma(\Upsilon(3S)\to \chi_{b1}(2P)+\gamma)=2.56\pm0.34$&$(1.57\pm0.10)\times10^{-1}$\\
     \hline
     $\Gamma(\psi^{\prime}\to \chi_{c2}+\gamma)=26.6\pm1.3$& $(2.74\pm0.07)\times10^{-1}$ &$\Gamma(\Upsilon(3S)\to \chi_{b2}(2P)+\gamma)=2.66\pm0.41$&$(1.55\pm0.12)\times10^{-1}$\\
     \hline
\end{tabular}
\end{table}

After including the spin-breaking contribution, the formula of the $E1$ transition decay widths turn to be:
\begin{subequations}
\begin{equation}
\Gamma(m^3S_1\to n ^3P_J)=(2J+1)\frac{(\delta_{QJ}^{nP,mS})^2}{9\pi}k_{\gamma}^{3}
\frac{M_{nP}}{M_{mS}}
\end{equation}
\begin{equation}
\Gamma(n^3P_J\to m ^3S_1)=\frac{(\delta_{QJ}^{nP,mS})^2}{3\pi}k_{\gamma}^{3}
\frac{M_{mS}}{M_{nP}},
\end{equation}
\begin{equation}
\Gamma(m ^1S_0\to n ^1P_1)=\frac{(\delta_{Q3}^{nP,mS})^2}{\pi}k_{\gamma}^{3}
\frac{M_{nP}}{M_{mS}}
\end{equation}
\begin{equation}
\Gamma(n ^1P_1\to m ^1S_0)=\frac{(\delta_{Q3}^{nP,mS})^2}{3\pi}k_{\gamma}^{3}
\frac{M_{mS}}{M_{nP}},
\end{equation}
\end{subequations}
where $k_{\gamma}$ is the energy of the emitted photon, and
\begin{subequations}\label{E1SP}
\begin{equation}
\delta_{Q0}^{nP,mS}=\delta_{Q}^{nP,mS}-\delta_{QS}^{nP,mS}+2\delta_{QL}^{nP,mS}+3\delta_{QT}^{nP,mS}
\end{equation}
\begin{equation}
\delta_{Q1}^{nP,mS}=\delta_{Q}^{nP,mS}-\delta_{QS}^{nP,mS}+\delta_{QL}^{nP,mS}-2\delta_{QT}^{nP,mS}
\end{equation}
\begin{equation}
\delta_{Q2}^{nP,mS}=\delta_{Q}^{nP,mS}-\delta_{QS}^{nP,mS}-\delta_{QL}^{nP,mS}+0\times\delta_{QT}^{nP,mS}
\end{equation}
\begin{equation}
\delta_{Q3}^{nP,mS}=\delta_{Q}^{nP,mS}+3\delta_{QS}^{nP,mS}-0\times\delta_{QL}^{nP,mS}+\delta_{QT}^{nP,mS}
\end{equation}
\end{subequations}

In principle, $\delta_{QJ}^{nP,mS}$ can be obtained by calculating the matrix element of
the electromagnetic current between the wave functions of the $n ^{3}P_{J}$ and $m ^{3}S_{1}$
states in pNRQCD (or in any potential model, see Ref. \cite{Eichten:2007qx} for a recent
review).
For some processes, such as $\chi_{cJ}\to J/\psi+\gamma$, $\psi^{\prime}\to \chi_{cJ}+\gamma$,
$\Upsilon(2S)\to \chi_{bJ}(1P)+\gamma$, and $\Upsilon(3S)\to \chi_{bJ}(2P)+\gamma$, their decay
widths have been measured \cite{Nakamura:2010px}, so we can obtain the values of the
corresponding coupling constants $\delta_{QJ}^{nP,mS}$ (for J=0,1,2) by fitting the data, which
are list in Table 1. Note, in our treatment of the uncertainties, we only take into account the
uncertainties in total decay widths and the branching ratios. The formulas in Eq.(\ref{E1SP}) show
that up to the sub-leading order $\delta_{QJ}^{nP,mS}$ (for J=0,1,2) only depends on
$\delta_{Q}^{nP,mS}-\delta_{QS}^{nP,mS}$, $\delta_{QL}^{nP,mS}$, and $\delta_{QT}^{nP,mS}$.
After resolving Eq.(\ref{E1SP}), we obtain that
\begin{eqnarray}\label{E1SP_n}
\delta_{c}^{1P,1S}-\delta_{cS}^{1P,1S}=22.7\times10^{-2}GeV^{-1},
\delta_{c}^{1P,2S}-\delta_{cS}^{1P,2S}=25.6\times10^{-2}GeV^{-1},\nonumber\\
\delta_{b}^{1P,2S}-\delta_{bS}^{1P,1S}=9.7\times10^{-2}GeV^{-1},
\delta_{b}^{2P,3S}-\delta_{bS}^{1P,2S}=15.3\times10^{-2}GeV^{-1},
\end{eqnarray}
The values of the corresponding $\delta_{QL}^{nP,mS}$, and $\delta_{QT}^{nP,mS}$ are given in
Table 2. If we assume that $\delta_{QS}^{nP,mS}$, $\delta_{QL}^{nP,mS}$, and $\delta_{QT}^{nP,mS}$
are in the same order, the results in Table 2 and those in Eq.(\ref{E1SP_n}) will indicate that
in these decay processes the contribution of the spin-breaking effect is less than that of the
leading order term by at least a factor of 10, and furthermore comparing to the $2S\to 1P$ transition
process in charmonium system the spin-breaking effect in the bottomonium system $2S\to 1P$ process
is of $(m_c/m_b)^2$ suppressed, which is consistent with the power counting rule of pNRQCD \cite{Pineda:2000sz}.


\begin{table}
\caption{The values of the coupling constants for the spin-breaking terms 
$\delta_{QL}^{nP,mS}$ and $\delta_{QT}^{nP,mS}$ determined by fitting the experimental 
decay widths $(\mathrm{unit} 10^{-2}\mathrm{GeV}^{-1})$.}
\begin{center}
\begin{tabular}{|c|c|c|c|c|c|c|c|}
     \hline
     $\delta_{cL}^{1p,1S}$&$\delta_{cT}^{1p,1S}$&$\delta_{cL}^{1p,2S}$&$\delta_{cT}^{1p,2S}$
     &$\delta_{bL}^{1p,2S}$&$\delta_{bT}^{1p,2S}$&$\delta_{bL}^{2p,3S}$&$\delta_{bT}^{2p,3S}$\\
     \hline
     $-0.3\pm0.4$&$-0.4\pm0.3$&$-1.8\pm0.3$&$0.1\pm0.2$&$-0.1\pm0.3$&$-0.1\pm0.2$&$-0.2\pm0.6$&$-0.3\pm0.3$\\
     \hline
\end{tabular}
\end{center}
\end{table}

\section{Radiative Transitions of $\psi(1^3D_{J^{\prime}})\to \chi_{cJ}+\gamma$}

The spectrum of $D$-wave heavy quarkonia has been calculated in potential model
by many groups, fox example recently in Ref.\cite{Barnes:2005pb,Godfrey:2001vc}.
In $c\bar{c}$ system, the $\psi(3770)$ state is treated as a pure $^3D_1$ state or
a predominant D-wave state with a small admixture of $2S$ state \cite{Ding:1991vu,Rosner:2001nm}.
The other states $^3D_2$, $^3D_3$ and $^1D_2$, whose decay widths are all expected
to be narrow, have not been observed yet. The masses of $\psi(^3D_2)$ and $\eta_{c2}(^1D_2)$
predicted by potential models lie between $D\bar{D}$ and $D\bar{D}^{\ast}$ thresholds
\cite{Barnes:2005pb}. They are forbidden to decay into pseudoscalar pair by parity.
The narrowness of $\psi(^3D_3)$ is due to that its decay into $D\bar{D}$ is a $F$-wave
decay, which is highly suppressed. Hence, the branching ratios of their radiative decays
are all considerable. The widths of the radiative transition $nD$ to $mP$ can be calculated
straightforwardly by employing the Lagrangian in Eq.(\ref{vertexPD}):
\begin{subequations}\label{E1PD}
\begin{eqnarray}\label{E1PD3}
\Gamma(n^3D_{J^{\prime}}\to m ^3P_J)=S_{J^{\prime},J}\frac{(\delta_{Q}^{nD,mP})^2}
{3\pi}k_{\gamma}^{3}\frac{M_{mP}}{M_{nD}}
\end{eqnarray}
\begin{eqnarray}\label{E1PD1}
\Gamma(n^1D_{2}\to m ^1P_1)=\frac{(\delta_{Q}^{nD,mP})^2}
{3\pi}k_{\gamma}^{3}\frac{M_{mP}}{M_{nD}},
\end{eqnarray}
\end{subequations}
where the coefficients $S_{J^{\prime},J}$ are $S_{1,J}=5/9,5/12,1/36$,
$S_{2,J}=0,3/4,1/4$, and $S_{3,J}=0,0,1$ for $J=0,1,2$ respectively \footnote{ For
the $^3D_{1}$ decay into $^3P_{J}+\gamma$, our results do not agree with those in
Ref.\cite{DeFazio:2008xq}. After private communications, their new results in the
erratum \cite{DeFazio1:2008xq} now agree with ours.}.

The decay widths of $\psi(3770)$ decay to $\chi_{c0}+\gamma$ and $\chi_{c1}+\gamma$
given in PDG are \cite{Nakamura:2010px}:
\begin{equation}\label{3D1}
\Gamma(\psi(3770)\to \chi_{c0}\gamma)=199\pm26\;\rm{keV},\;
\Gamma(\psi(3770)\to \chi_{c1}\gamma)=79\pm17\rm{keV}.
\end{equation}

If $\psi(3770)$ is a pure $1D$ state, the values of the coupling constant
$\delta_{c}^{1D1P}$ determined through $\Gamma(\psi(3770)\to \chi_{cJ}+\gamma$ are
$\delta_{c}^{1D1P}=0.31\pm0.02\;\rm{GeV}^{-1}$, and $\delta_{c}^{1D1P}=0.35\pm0.04\;\rm{GeV}^{-1}$
for $J=0$ and $J=1$, respectively, which are very close to each other. In this case, it
indicates that the heavy quark spin symmetry is roughly preserved in the $1D$ to $1P$
radiative transitions. The average value is
\begin{equation}\label{3D10}
\bar{\delta}_{c}^{1D1P}=0.32\pm0.02\;\rm{GeV}^{-1}.
\end{equation}

If we treat $\psi(3770)$ as a S-D mixing state and using the same notation in Ref.\cite{Rosner:2001nm}:
\begin{equation}\label{SDmxing}
\psi(3686)=\cos(\theta)|2S\rangle-\sin(\theta)|1D\rangle,\;
\psi(3770)=\cos(\theta)|1D\rangle+\sin(\theta)|2S\rangle,
\end{equation}
the analytical formulas for $\psi(3686)$ and $\psi(3770)$ decay into $\gamma+\chi_{cJ}$ turn to
be
\begin{subequations}
\begin{equation}
\Gamma(\psi(3686)\to \chi_{c0}+\gamma)=\frac{M_{\chi_{c0}}}{9\pi M_{\psi(3686)}}k_{\gamma}^{3}\,
(\cos^2(\theta)(\delta_{c}^{1P,2S})^2
-2\sqrt{\frac{5}{3}}\sin(\theta)\cos(\theta)\delta_{c}^{1P,2S}\delta_{c}^{1D,1P}
+\frac{5}{3}\sin^2(\theta)(\delta_{c}^{1D,1P})^2)
\end{equation}
\begin{equation}
\Gamma(\psi(3686)\to \chi_{c1}+\gamma)=\frac{3M_{\chi_{c1}}}{9\pi M_{\psi(3686)}}k_{\gamma}^{3}\,
(\cos^2(\theta)(\delta_{c}^{1P,2S})^2
+\sqrt{\frac{5}{3}}\sin(\theta)\cos(\theta)\delta_{c}^{1P,1S}\delta_{c}^{1D,1P}
+\frac{5}{12}\sin^2(\theta)(\delta_{c}^{1D,1P})^2)
\end{equation}
\begin{equation}
\Gamma(\psi(3686)\to \chi_{c2}+\gamma)=\frac{5M_{\chi_{c2}}}{9\pi M_{\psi(3686)}}k_{\gamma}^{3}\,
(\cos^2(\theta)(\delta_{c}^{1P,2S})^2
-\sqrt{\frac{1}{15}}\sin(\theta)\cos(\theta)\delta_{c}^{1P,2S}\delta_{c}^{1D,1P}
+\frac{1}{60}\sin^2(\theta)(\delta_{c}^{1D,1P})^2)
\end{equation}
\begin{equation}
\Gamma(\psi(3770)\to \chi_{c0}+\gamma)=\frac{5M_{\chi_{c0}}}{27\pi M_{\psi(3770)}}k_{\gamma}^{3}\,
(\cos^2(\theta)(\delta_{c}^{1D,1P})^2
+2\sqrt{\frac{3}{5}}\sin(\theta)\cos(\theta)\delta_{c}^{1D,1P}\delta_{c}^{1P,2S}
+\frac{3}{5}\sin^2(\theta)(\delta_{c}^{1P,2S})^2)
\end{equation}
\begin{equation}
\Gamma(\psi(3770)\to \chi_{c1}+\gamma)=\frac{5M_{\chi_{c1}}}{36\pi M_{\psi(3770)}}k_{\gamma}^{3}\,
(\cos^2(\theta)(\delta_{c}^{1D,1P})^2
-4\sqrt{\frac{3}{5}}\sin(\theta)\cos(\theta)\delta_{c}^{1D,1P}\delta_{c}^{1P,2S}
+\frac{12}{5}\sin^2(\theta)(\delta_{c}^{1P,2S})^2)
\end{equation}
\begin{equation}
\Gamma(\psi(3770)\to \chi_{c2}+\gamma)=\frac{M_{\chi_{c2}}}{108\pi M_{\psi(3770)}}k_{\gamma}^{3}\,
(\cos^2(\theta)(\delta_{c}^{1D,1P})^2
+4\sqrt{15}\sin(\theta)\cos(\theta)\delta_{c}^{1D,1P}\delta_{c}^{1P,2S}
+60\sin^2(\theta)(\delta_{c}^{1P,2S})^2)
\end{equation}
\end{subequations}
The mixing angle $\theta=(12\pm2)^{\circ}$ that is determined from the leptonic decays of
$\psi(3686)$ and $\psi(3770)$ also is favored by some other considerations\cite{Kuang:1989ub}.
If we fit $\psi(3770)$ and $(\psi(3686))$ decays into $\chi_{c0}+\gamma$ with $\theta=12^{\circ}$,
we obtain two set solutions, which are labeled by subscripts $1$ and $2$, respectively,
\begin{equation}\label{SD0}
\delta_{1}^{1P,2S}=0.30\rm{GeV}^{-1},\delta_{1}^{1D,1P}=0.27\rm{GeV}^{-1};\;
\delta_{2}^{1P,2S}=0.14\rm{GeV}^{-1},\delta_{2}^{1D,1P}=-0.34\rm{GeV}^{-1};
\end{equation}
If we fit their decay into $\chi_{c1}+\gamma$ the results are:
\begin{equation}\label{SD1}
\delta_{1}^{1P,2S}=0.18\rm{GeV}^{-1},\delta_{1}^{1D,1P}=0.42\rm{GeV}^{-1};\;
\delta_{2}^{1P,2S}=0.28\rm{GeV}^{-1},\delta_{2}^{1D,1P}=-0.27\rm{GeV}^{-1};
\end{equation}
The difference between the results in Eq.(\ref{SD0}) and those in Eq.(\ref{SD1}) shows
that the heavy quark spin symmetry is largely violated if $\psi(3686)$ and $\psi(3770)$
are assigned as two S-D mixing sates as given in Eq.(\ref{SDmxing}) with mixing angel
$\theta=(12\pm2)^{\circ}$. Consequently, to understand the radiative decays of $\psi(3770)$
and $\psi(3686)$ in the $S-D$ mixing picture, some other effects like the relativistic
corrections\cite{Li:2009zu} or the couple channels effect\cite{Eichten:2004uh} should
be taken into account.

Since in the S-D mixing picture the heavy quark spin symmetry does not hold anymore,
thereafter we will adopt that $\psi(3770)$ is a pure D-wave state and choose the value
of the coupling constant to be that in Eq.(\ref{3D10}) to study the radiative decay of
the D-wave states. The upper limit of $\psi(3770)\to \chi_{c2}+\gamma$ is that
$\mathcal{B}(\psi(3770)\to \chi_{c2}+\gamma)<9\times10^{-4}$ \cite{Nakamura:2010px}.
Using the result in Eq.(\ref{E1PD3}), we predict that
\begin{equation}
\Gamma(\psi(3770)\to \chi_{c2}+\gamma)=2.55\pm0.28\;\rm{keV},\;
\mathcal{B}(\psi(3770)\to \chi_{c2}+\gamma)=(9.4\pm1.0)\times10^{-5}.
\end{equation}
which is compatible with the experimental data and is about 4 times larger than those in
Ref.\cite{DeFazio:2008xq}. The $\psi(3770)$ decay into $\chi_{cJ}+\gamma$ has also been
studied by potential model. For comparison, we choose two potential models calculations
\cite{Ding:1991vu,Barnes:2005pb}, in which the predictions of $\psi(3770)$ decay into
$\chi_{c0,1}+\gamma$ agree well with the experimental data after including
relativistic corrections. Their predictions of $\Gamma(\psi(3770)\to \chi_{c2}+\gamma)$
are $3.0$\cite{Ding:1991vu} or $3.3$\cite{Barnes:2005pb} keV. Both of them are consistent
with our results.

As mentioned above, the other $1D$ states are all expected to be narrow. Their spectrum
and the E1 transition decay widths have also been calculated in Ref.\cite{Barnes:2005pb}.
Their results are:
\begin{subequations}\label{E11D1P}
\begin{equation}
M(1^3D_2)=3.838\rm{GeV},\;\Gamma(1^3D_2\to \chi_{c1}(\chi_{c2})+\gamma)=268(66)\;\rm{keV}
\end{equation}
\begin{equation}
M(1^3D_3)=3.849\rm{GeV},\;\Gamma(1^3D_3\to \chi_{c2}+\gamma)=296\;\rm{keV}
\end{equation}
\begin{equation}
M(1^1D_2)=3.837\rm{GeV},\;\Gamma(1^1D_2\to h_{c}+\gamma)=344\;\rm{keV}.
\end{equation}
\end{subequations}
If we choose the same mass values, our predictions are
\begin{subequations}
\begin{equation}
\Gamma(1^3D_2\to \chi_{c1}+\gamma)=(288\pm25)\;\rm{keV},\;
\Gamma(1^3D_2\to \chi_{c2}+\gamma)=(50.3\pm5.5)\;\rm{keV},
\end{equation}
\begin{equation}
\Gamma(1^3D_3\to \chi_{c2}+\gamma)=224\pm25\;\rm{keV},\;
\Gamma(1^1D_2\to h_{c}+\gamma)=267\pm29\;\rm{keV}.
\end{equation}
\end{subequations}
which agree with the potential model results.

Recently, the $X(3872)$ state has received much attention since it was first discovered by
Belle Collaboration \cite{Choi:2003ue}, and then was confirmed in $p\bar{p}$ collision at
Tevatron \cite{Acosta:2003zx}. It was also observed  by Babar Collaboration \cite{Aubert:2004ns}.
Until now, there is not a convincing explanation about its nature yet. Only the charge parity
$C=+$ is established from its decay into $J/\psi+\gamma$ \cite{Abe:2005ix}. After analyzing
$B\to J/\psi+\omega+K$, Babar Collaboration found its $J^{PC}$ favors $2^{-+}$ \cite{delAmoSanchez:2010jr}.
If it is a pure charmonium D-wave state, the only assignment will be the $\eta_{c2}(^1D_2)$.
It then should has a sizeable decay into $\gamma+h_c$. Evaluating in a similar way, we obtain
\begin{equation}
\Gamma(X(3872)\to h_c+\gamma)=359\pm59\rm{keV},
\end{equation}
which is very large. So studying $X(3872)$ decay into $\gamma+h_c$ will be helpful to understand
its nature.

The $D$-wave bottomonium states were observed from the cascade of $\Upsilon(3S)$ \cite{Bonvicini:2004yj},
however no other further information is known yet. We can not make any prediction about their
radiative decay with the effective Lagrangian method at present.

\section{Relation Between Radiative Decay and LH Decay of $\chi_{bJ}(nP)$}
The total decay widths of the P-wave bottomonium states $\chi_{bJ}(nP)\; (n=1,2)$ have not
been measured yet, so we can not compute $\delta_{b}^{nP,mS}$ by fitting the data. Besides
the radiative decay, the LH decay is also an important decay mode for the P-wave quarkonium
states. One remarkable success of NRQCD is that it can systematically resolve the infrared
divergence problem in the CS model (CSM) calculation for the LH decays of $P$-wave states
by introducing the CO contribution \cite{Bodwin:1994jh,Bodwin:1992ye}. For the states in
strong coupling region, where most of the heavy quarkonium states below threshold are expected
to belong to, further study of pNRQCD shows that the CO matrix elements can be related to the
wave function of the bound states \cite{Brambilla:2001xy,Brambilla:2002nu}. In particular,
in the strong coupling region the ratio $\rho_{8}(nP)=m_b^2\mathcal{H}_{8}(nP)/\mathcal{H}_{1}(nP)$
does not dependent very much on the radial quantum number $n$ \cite{Brambilla:2001xy}.
By fitting the open charm decays of $\chi_{bJ}(1P)$ and $\chi_{bJ}(2P)$, CLEO collaboration
obtained that $\rho_{8}(1P)=0.160^{+0.071}_{-0.047}$ and $\rho_{8}(2P)=0.074^{+0.010}_{-0.008}$
\cite{Briere:2008cv}, which is a little different from pNRQCD prediction.

Next, we will show that the relation between the radiative decay and the LH decay could provide
another way to extract the values of $\rho_{8}(1,2P)$. According to NRQCD approach, at $v_b$
leading order, the LH decay width for P-wave states is given by:
\begin{equation}
\Gamma( \chi_{bJ}(nP)\to LH )=\frac{C_1(\mu)\mathcal{H}_{1}}{m_b^4}+\frac{C_{8}(\mu)\mathcal{H}_{8}}{m_b^2}
\end{equation}
where $C_1$ and $C_8$ are the $J$-dependent short distance coefficients and have been calculated
up to $\alpha_s^{3}$ order \cite{Petrelli:1997ge}. Although neither of the radiative and LH decay
widths have been measured, their branching ratios are known. Recently Babar Collaboration update
the branching ratios of $\chi_{bJ}(nP)\to \Upsilon(mS)+\gamma$, their latest results are \cite{Lees:2011mx}:
\begin{subequations}
\begin{equation}
\mathcal{B}\chi_{bJ}(1P) \to \Upsilon(1S) +\gamma= (2.2\pm1.5^{+1.0}_{-0.7}\pm0.2,34.9\pm0.8
\pm2.2\pm2.0,19.5\pm0.7^{+1.3}_{-1.5}\pm1.0)\%\; \rm{for} \;(J=0,1,2)
\end{equation}
\begin{equation}
\mathcal{B}\chi_{bJ}(2P) \to \Upsilon(2S) + \gamma=
(-4.7\pm2.8^{+0.7}_{-0.8}\pm0.5,18.9\pm1.1\pm1.2\pm1.8,8.3\pm0.8\pm0.6\pm1.0)\%\; \rm{for} \;(J=0,1,2)
\end{equation}
\begin{equation}
\mathcal{B}\chi_{bJ}(2P) \to \Upsilon(1S) + \gamma=
(0.7\pm0.4^{+0.2}_{-0.1}\pm0.1,9.9\pm0.3^{+0.5}_{-0.4}\pm0.9,7.0\pm0.2\pm0.3\pm0.9)\%\; \rm{for} \;(J=0,1,2)
\end{equation}
\end{subequations}
The branching ratio of $\chi_{bJ}(nP)$ decay into LH can be obtained by subtracting its all
the known transitions to other bottomonium states, which can be read out directly from PDG
\cite{Nakamura:2010px}. The ratio of the two branching ratios can be expressed as
\begin{equation}\label{E1LH}
R_{J}(nP)=\frac{\mathcal{B}( \chi_{bJ}(nP)\to LH )}{\mathcal{B}( \chi_{bJ}(nP)\to \gamma +\Upsilon(nS))}
=\frac{\Gamma( \chi_{bJ}(nP)\to LH )}{\Gamma( \chi_{bJ}(nP)\to \gamma +\Upsilon(nS))}
=\frac{3\pi M_{nP}}{M_{nS}\delta^{nP,nS}k_{\gamma}^3}(C_1(\mu)\mathcal{H}_{1}(nP)+C_{8}(\mu)\mathcal{H}_{8}(nP))
\end{equation}
$R_{J}(nP)$ only dependents on three unknown parameters $\mathcal{H}_{1}(nP)$, $\mathcal{H}_{8}(nP)$ and
$\delta_{b}^{nP,nS}$. Therefore, to compute the ratio $\rho_{8}(nP)$, we only need two independent inputs.
Since the uncertainties of the $\chi_{b0}(1,2P)$ radiative decays are are large, we choose the data of
$\chi_{b1,2}(1,2P)$ decays. Using the $\alpha_{s}^{3}$ order short-distances coefficients listed in Ref.
\cite{Petrelli:1997ge} and setting $\mu_{R}=\mu=2m_b$, $\alpha_s(2m_b)=0.18$, and the number of light
flavor quark $N_f=4$, we obtain $\rho_{8}(1P)=0.150^{+0.036}_{-0.037}$ and $\rho_{8}(2P)=0.110\pm0.030$.
Our value of $\rho_{8}(1P)$ is a little smaller than that of CLEO, while our value of $\rho_{8}(2P)$
is about 1.5 times larger than that of CLEO, which makes $\rho_{8}(1P)$ close to $\rho_{8}(2P)$. This
indicates that the pNRQCD assumptions is reasonable to study the LH decays of $\chi_{bJ}(1P)$ and
$\chi_{bJ}(2P)$ states \cite{Brambilla:2001xy}.

\section{Summary and Conclusion}
In summary, the radiative decays of the heavy quarkonia are studied with the help of the effective
Lagrangian. To have a better understanding of the radiative transitions among $S-$ and $P-$ wave
states, we take into account the contribution that is due to the spin-breaking interactions. By
fitting the experimental data, the coupling constants of the spin-breaking terms in $\psi(2S)$,
$\chi_{cJ}$, $\Upsilon(2S)$ and $\Upsilon(3S)$ radiative are obtained, which are listed in Table 2.
We find that the values of the coupling constants in the spin-breaking terms are less than those in
the leading order terms by at least a factor of 10 and that the spin-breaking terms in $\psi(2S)$
and $\Upsilon(2S)$ indicate that the spin-breaking contribution is suppressed by $1/m_Q^2$, which
agrees with pNRQCD power counting rule. We also calculate the radiative decays of the $c\bar{c}(1D)$
states, whose total decay widths are expected to be narrow. Based on that $\psi(3770)$ is a pure
$D$-wave state, our predictions of the radiative decay widths of the other $D$-wave states are
consistent with the potential model results. Furthermore, we predict that
$\Gamma(X(3872)\to h_c+\gamma)=359\pm59\rm{keV}$, if $X(3872)$ is the $2^{-+}$ state. We also study
the S-D mixing effect in $\psi(3770)$ radiative decay and find that there is no heavy quark spin-symmetry
if the mixing angel is $12^{\circ}$.  As an useful application, we find that relating the radiative
decay of $\chi_{bJ}(nP)$ to their LH decays can provide another way to estimate the ratios of
$\rho_{8}(nP)=m_b^2\mathcal{H}_{8}(nP)/\mathcal{H}_{1}(nP)$. By fitting the data of $\chi_{b1}(1P,2P)$
and that of $\chi_{b2}(1P,2P)$, we get that $\rho_{8}(1P)=0.150^{+0.036}_{-0.037}$ and $\rho_{8}(2P)=0.110\pm0.030$,
which approximately equal to each other. Our result provide an evidence on pNRQCD assumptions
\cite{Brambilla:2001xy}.

\section{Acknowledgement}
The author is grateful to Prof.Joan Soto for very helpful discussions and carefully reading the manuscript.
This work is supported by the CSD2007-00042 Consolider-Ingenio 2010 program under Contract No. CPAN08-PD14,
and by the FPA2007-66665-C02-01/ and FPA2010-16963 projects (Spain).







\begin{thebibliography}{00}


\bibitem{Swanson:2006st}
  E.~S.~Swanson,
  Phys.\ Rept.\  {\bf 429}, 243 (2006);
  S.~L.~Zhu,
  Int.\ J.\ Mod.\ Phys.\  E {\bf 17}, 283 (2008);
  N.~Brambilla {\it et al.},
  Eur.\ Phys.\ J.\  C {\bf 71}, 1534 (2011).


\bibitem{Caswell:1985ui}
  W.~E.~Caswell and G.~P.~Lepage,
  Phys.\ Lett.\  B {\bf 167}, 437 (1986).


\bibitem{Thacker:1990bm}
  B.~A.~Thacker and G.~P.~Lepage,
  Phys.\ Rev.\  D {\bf 43}, 196 (1991).


\bibitem{Bodwin:1994jh}
  G.~T.~Bodwin, E.~Braaten and G.~P.~Lepage,
  Phys.\ Rev.\  D {\bf 51}, 1125 (1995)
  [Erratum-ibid.\  D {\bf 55}, 5853 (1997)].


\bibitem{Pineda:1997bj}
  A.~Pineda and J.~Soto,
  Nucl.\ Phys.\ Proc.\ Suppl.\  {\bf 64}, 428 (1998).


\bibitem{Brambilla:1999xf}
  N.~Brambilla, A.~Pineda, J.~Soto and A.~Vairo,
  Nucl.\ Phys.\  B {\bf 566}, 275 (2000).


\bibitem{Brambilla:2004jw}
  N.~Brambilla, A.~Pineda, J.~Soto and A.~Vairo,
  Rev.\ Mod.\ Phys.\  {\bf 77}, 1423 (2005).


\bibitem{Brambilla:2005zw}
  N.~Brambilla, Y.~Jia and A.~Vairo,
  Phys.\ Rev.\  D {\bf 73}, 054005 (2006);
  Y.~Jia, W.~L.~Sang and J.~Xu,
  arXiv:1007.4541 [hep-ph].




\bibitem{Eichten:2007qx}
  E.~Eichten, S.~Godfrey, H.~Mahlke and J.~L.~Rosner,
  Rev.\ Mod.\ Phys.\  {\bf 80}, 1161 (2008).





\bibitem{Brambilla:2001xy}
  N.~Brambilla, D.~Eiras, A.~Pineda, J.~Soto and A.~Vairo,
  Phys.\ Rev.\ Lett.\  {\bf 88}, 012003 (2002).


\bibitem{Brambilla:2002nu}
  N.~Brambilla, D.~Eiras, A.~Pineda, J.~Soto and A.~Vairo,
  Phys.\ Rev.\  D {\bf 67}, 034018 (2003).


\bibitem{Neubert:1993mb}
  M.~Neubert,
  Phys.\ Rept.\  {\bf 245}, 259 (1994).


\bibitem{Casalbuoni:1992yd}
  R.~Casalbuoni, A.~Deandrea, N.~Di Bartolomeo, R.~Gatto, F.~Feruglio and G.~Nardulli,
  Phys.\ Lett.\  B {\bf 302}, 95 (1993).


\bibitem{DeFazio:2008xq}
  F.~De Fazio,
  Phys.\ Rev.\  D {\bf 79}, 054015 (2009).


\bibitem{Pineda:2000sz}
  A.~Pineda and A.~Vairo,
  Phys.\ Rev.\  D {\bf 63}, 054007 (2001);
  [Erratum-ibid.\  D {\bf 64}, 039902 (2001)].


\bibitem{Nakamura:2010px}
  K. Nakamura {\it et al.}  [Particle Data Group],
  J.\ Phys.\ G {\bf 37}, 075021 (2010).


\bibitem{Barnes:2005pb}
  T.~Barnes, S.~Godfrey and E.~S.~Swanson,
  Phys.\ Rev.\  D {\bf 72}, 054026 (2005);

\bibitem{Godfrey:2001vc}
  S.~Godfrey and J.~L.~Rosner,
  Phys.\ Rev.\  D {\bf 64}, 097501 (2001)
  [Erratum-ibid.\  D {\bf 66}, 059902 (2002)].


\bibitem{Ding:1991vu}
  Y.~B.~Ding, D.~H.~Qin and K.~T.~Chao,
  Phys.\ Rev.\  D {\bf 44}, 3562 (1991).


\bibitem{Rosner:2001nm}
  J.~L.~Rosner,
  Phys.\ Rev.\  D {\bf 64}, 094002 (2001).


\bibitem{DeFazio1:2008xq}
  F.~De Fazio,
 Erratum-ibid.\  D {\bf 83}, 099901 (2011).


\bibitem{Kuang:1989ub}
  Y.~P.~Kuang and T.~M.~Yan,
  Phys.\ Rev.\  D {\bf 41}, 155 (1990).


\bibitem{Li:2009zu}
  B.~Q.~Li and K.~T.~Chao,
  Phys.\ Rev.\  D {\bf 79}, 094004 (2009).


\bibitem{Eichten:2004uh}
  E.~J.~Eichten, K.~Lane and C.~Quigg,
  Phys.\ Rev.\  D {\bf 69}, 094019 (2004);
  E.~J.~Eichten, K.~Lane and C.~Quigg,
  Phys.\ Rev.\  D {\bf 73}, 014014 (2006)
  [Erratum-ibid.\  D {\bf 73}, 079903 (2006)].





\bibitem{Choi:2003ue}
  S.~K.~Choi {\it et al.}  [Belle Collaboration],
  Phys.\ Rev.\ Lett.\  {\bf 91}, 262001 (2003).


\bibitem{Acosta:2003zx}
  D.~E.~Acosta {\it et al.}  [CDF II Collaboration],
  Phys.\ Rev.\ Lett.\  {\bf 93}, 072001 (2004);
  V.~M.~Abazov {\it et al.}  [D0 Collaboration],
  Phys.\ Rev.\ Lett.\  {\bf 93}, 162002 (2004).


\bibitem{Aubert:2004ns}
  B.~Aubert {\it et al.}  [BABAR Collaboration],
  Phys.\ Rev.\  D {\bf 71}, 071103 (2005).


\bibitem{Abe:2005ix}
  K.~Abe {\it et al.}  [Belle Collaboration],
  arXiv:hep-ex/0505037.


\bibitem{delAmoSanchez:2010jr}
  P.~del Amo Sanchez {\it et al.}  [BABAR Collaboration],
  Phys.\ Rev.\  D {\bf 82}, 011101 (2010).


\bibitem{Bonvicini:2004yj}
  G.~Bonvicini {\it et al.}  [CLEO Collaboration],
  Phys.\ Rev.\  D {\bf 70}, 032001 (2004);
  P.~del Amo Sanchez {\it et al.}  [BABAR Collaboration],
  Phys.\ Rev.\  D {\bf 82}, 111102 (2010).


\bibitem{Bodwin:1992ye}
  G.~T.~Bodwin, E.~Braaten and G.~P.~Lepage,
  Phys.\ Rev.\  D {\bf 46}, 1914 (1992).


\bibitem{Briere:2008cv}
  R.~A.~Briere {\it et al.}  [CLEO Collaboration],
  Phys.\ Rev.\  D {\bf 78}, 092007 (2008).


\bibitem{Lees:2011mx}
  J.~P.~Lees {\it et al.}  [The BABAR Collaboration],
  Phys.\ Rev.\  D {\bf 84}, 072002 (2011)
  [Phys.\ Rev.\  D {\bf 84}, 099901 (2011)]
  [arXiv:1104.5254 [hep-ex]].



\bibitem{Petrelli:1997ge}
  A.~Petrelli, M.~Cacciari, M.~Greco, F.~Maltoni and M.~L.~Mangano,
  Nucl.\ Phys.\  B {\bf 514}, 245 (1998).

\end{thebibliography}



\end{document}